\documentclass[10pt,journal,compsoc]{IEEEtran}
\ifCLASSOPTIONcompsoc
  \usepackage[nocompress]{cite}
\else
  \usepackage{cite}
\fi
\ifCLASSINFOpdf
   \usepackage[pdftex]{graphicx}
  \graphicspath{{../pdf/}{../jpeg/}}
  \DeclareGraphicsExtensions{.pdf,.jpeg,.png}
\else
  \usepackage[dvips]{graphicx}
  \graphicspath{{../eps/}}
  \DeclareGraphicsExtensions{.eps}
\fi
\usepackage{amsmath}
\usepackage{cases}

\usepackage{algorithm}
\usepackage{algorithmic}

\newcommand{\BEGIN}[1]{\STATE \textbf{Begin} \begin{ALC@g}} 
\newcommand{\END}{\end{ALC@g} \STATE \textbf{End}}
\newcommand{\SWITCH}[1]{\STATE \textbf{switch} (#1)}
\newcommand{\ENDSWITCH}{\STATE \textbf{end switch}}
\newcommand{\CASE}[1]{\STATE \textbf{case} #1\textbf{:} \begin{ALC@g}}
\newcommand{\ENDCASE}{\end{ALC@g}}

\newcommand{\DEFAULT}{\STATE \textbf{default:} \begin{ALC@g}}
\newcommand{\ENDDEFAULT}{\end{ALC@g}}
\newcommand{\DEFAULTLINE}[1]{\STATE \textbf{default:} }
\newcommand{\ON}[1]{\STATE \textbf{On} #1\textbf{:} \begin{ALC@g}} 
\newcommand{\ENDON}{\end{ALC@g}}

\ifCLASSOPTIONcompsoc
  \usepackage[caption=false,font=footnotesize,labelfont=sf,textfont=sf]{subfig}
\else
  \usepackage[caption=false,font=footnotesize]{subfig}
\fi
\usepackage{url}
\usepackage[outdir=./]{epstopdf}

\begin{document}
%
\title{Nik Defense: An Artificial Intelligence Based Defense Mechanism against Selfish Mining in Bitcoin}
%
%
%
%

\author{Ali~Nikhalat-Jahromi,
        Ali~Mohammad~Saghiri,
        and~Mohammad~Reza~Meybodi
\IEEEcompsocitemizethanks{\IEEEcompsocthanksitem Ali Nikhalat Jahromi, Ali Mohammad Saghiri, and Mohammad Reza Meybodi are with the Department
of Computer Engineering, Amirkabir University of Technology, Tehran,
Iran.\protect\\
E-mail: \{ali.nikhalat,a\_m\_saghiri,mmeybodi\}@aut.ac.ir}
}

\IEEEtitleabstractindextext{%
\begin{abstract}
The Bitcoin cryptocurrency has received much attention recently. In the network of Bitcoin, transactions are recorded in a ledger. In this network, the process of recording transactions depends on some nodes called miners that execute a protocol known as mining protocol. One of the significant aspects of mining protocol is incentive compatibility. However, literature has shown that Bitcoin mining's protocol is not incentive-compatible. Some nodes with high computational power can obtain more revenue than their fair share by adopting a type of attack called the selfish mining attack. In this paper, we propose an artificial intelligence-based defense against selfish mining attacks by applying the theory of learning automata. The proposed defense mechanism ignores private blocks by assigning weight based on block discovery time and changes current Bitcoin's fork resolving policy by evaluating branches' height difference in a self-adaptive manner utilizing learning automata. To the best of our knowledge, the proposed protocol is the literature's first learning-based defense mechanism. Simulation results have shown the superiority of the proposed mechanism against tie-breaking mechanism, which is a well-known defense. The simulation results have shown that the suggested defense mechanism increases the profit threshold up to 40\% and decreases the revenue of selfish attackers.

\end{abstract}

}

\maketitle

\IEEEdisplaynontitleabstractindextext

%
\IEEEpeerreviewmaketitle

\IEEEraisesectionheading{\section{Introduction}\label{sec:introduction}}

%
%
%
%
\IEEEPARstart{B}{itcoin} \cite{RN171}, a decentralized cryptocurrency, was introduced by Satoshi Nakamoto in 2009 \cite{RN172, RN173, RN181, RN177}. It has attracted much attention because of implementing a fully trustable decentralized financial system. In Bitcoin network, manipulating financial transactions is done using blockchain technology. In this technology, the system records all transactions between Bitcoin clients. The security of the blockchain depends on a cryptographic puzzle. Some of the participants in the blockchain's network, called miners, try to solve a cryptographic puzzle for putting transactions in the newly discovered block. Each miner hopes to put the newly discovered block into the main chain to obtain a reward. The mining process is incentive-compatible, meaning that a miner who solves the cryptographic puzzle gets a reward based on resource sharing in the mining process \cite{RN179, RN201, RN191, RN193, RN197}. In the next paragraph, the mining process and its challenges are highlighted.

In the mining process \cite{RN179, RN201, RN191, RN193, RN197}, miners compete collectively to discover and broadcast new blocks, but sometimes more than one block is ahead of the preceding block. A fork has been created on top of the chain in this situation. To resolve forks, the protocol suggests adopting and mining on the longest chain, which has the chain with the most work, or in the situation with the same chain length, the miner should choose the first received block. The following two paragraphs explain the challenges of the mining protocol of Bitcoin, which leads to selfish mining attacks.

The main drawback of Bitcoin protocols is that the system works under some assumptions that are not true in all situations \cite{RN199, RN183, RN175}, as explained as follows. The Bitcoin protocols require more than half of miners to be honest, which means that they should follow the mining process without any changes, But Eyal and Sirer \cite{RN184} have shown that this assumption might not be correct in some situations. In their work, they introduced a new mining strategy, denoting the selfish mining strategy version 1, which can be abbreviated as SM1; In the SM1's strategy, miners try to selfishly increase their revenue by keeping newly discovered blocks private and creating a new fork. Honest miners continue to mine on the public chain, while selfish miners continue to mine on the private chain they started. If selfish miners discover more blocks than the others, they will try to keep newly discovered blocks private. This effort aims to develop a more extended chain basis on the current public chain. When the public chain approaches the selfish miners' private chain in length, they will reveal blocks from their private chain to the public
\cite{RN184, RN186}.

The selfish mining attack might threaten the decentralization and fairness capabilities of the bitcoin network. By increasing the mining power threshold, selfish miners can obtain more revenue than their fair share. Also, honest miners persuade to leave Bitcoin's honest mining protocol. Honest miners prefer to join selfish miners by following selfish mining protocols; joining selfish miners causes them to increase selfish mining power quickly. If selfish miners' computational power reaches the majority threshold, it can lead to a new protocol that discards other miners' discovered blocks.

In this paper, we propose the Nik defense mechanism, the first artificial intelligence-based defense against selfish mining based on one of the reinforcement learning methods called learning automata. The proposed defense mechanism manipulates the fork-resolving policy, leading to a novel resolving approach. In the proposed mechanism, for the first time in the literature, a defense algorithm based on learning automata theory and a novel weighting mechanism are suggested for managing operations on the chain in the blockchain network in a self-organized manner. In other words, the system can reorganize itself against selfish mining attacks to decrease the profit of selfish miners. In addition, another policy will be suggested to change the fail-safe parameter by learning automaton on every miner in the blockchain. This change in the fail-safe parameter also makes it difficult for selfish miners to decide between publishing or keeping newly discovered blocks.

To show the effectiveness of a new proposed defense, we compare the proposed defense with the most famous defense, called the tie-breaking defense \cite{RN184}. The rest of the paper is organized as follows. The preliminaries related to the selfish mining attack are given in section 2. In section 3, related work is being discussed. In section 4, we describe the proposed defense mechanism. In section 5, the modified version of SM1 for evaluation is discussed. Section 6 reports the result of the experiments, and section 7, discusses the paper's limitations and problems. Finally, section 8 concludes the paper.
    
\section{Preliminaries}
In this section, required information about the proposed algorithm is given. We summarize information about blockchain, selfish mining strategies, and learning automata. In the rest of this section, at first, an overview of the Bitcoin and its transaction is given. Then the mining process, selfish mining attack, learning automata, and properties of an ideal defense are explained, respectively.

Bitcoin is a distributed and decentralized cryptocurrency. The users of Bitcoin can transfer Bitcoins by creating a new transaction \cite{RN203, RN205} and sending it to a ledger based on the blockchain. The blockchain is an append-only ledger protected by a group of miners in the network. Miners are rewarded for their effort to protect the blockchain against tampering data. A transaction in the blockchain is made of at least one input and one output. The difference between the total amount of inputs and outputs in a transaction is called a transaction fee \cite{RN207, RN209}. The transaction fee goes to the miner, who includes the transaction in the blockchain. The following subsection gives more details about the mining process.

\subsection{Mining Process}
The state of the blockchain is changed through transactions \cite{RN213}. Transactions are grouped into blocks that are appended to the blockchain. A typical block in blockchain consists of two major parts: header and body. The block's header contains the hash of the previous block, the hash of the current block, the Merkle root of all transactions included in this block, and a number called the nonce. The block's body contains transactions that the miner decided to include in the block \cite{RN172, RN173}.

A valid block contains a solution to a puzzle. To solve the puzzle, miners try to put the correct nonce in the block's header so that the block's hash is smaller than the block difficulty target \cite{RN213, RN215}. The block difficulty is dynamically adjusted such that blocks are generated at an average rate of one every ten minutes. If a miner solves the puzzle and puts mined block in the longest chain, it will be rewarded with Bitcoins that did not exist before and the transaction fees of the newly created block.

The probability of mining a new block is proportional to the computational resources used for solving the associated puzzle. Due to the nature of the mining process, the interval between mining events exhibits high variance from the point of view of a single miner. Consequently, miners typically organize themselves into mining pools. All pool members work together to mine each block and share their revenues when one of them successfully mines a block. While joining a pool does not change a miner's expected revenue, it decreases the variance and makes the monthly revenue more predictable \cite{RN184}. 

\subsection{Selfish Mining Attacks}
Bitcoin's doc illustrates the approach of releasing blocks after mining. A miner who discovers a new valid block should release it immediately. Eyal and Sirer \cite{RN184} showed that some miners could gain revenue more than their fair share by deviating from Bitcoin's main mining rules, which they called "selfish mining." As noted, miners try to form a pool to decrease revenue variance. So a selfish pool with more than 1/3 computational power unfairly changes Bitcoin's rewarding system by keeping a private chain and withholding blocks that have been mined.

Saphirshtein et al. \cite{RN216} used Markov Decision Process (MDP) to investigate the profit threshold (the minimal fraction of resources required for a profitable attack). They find the bound under which the system can be considered secure against such attacks and modify the protocol to assess their susceptibility to selfish mining by computing the optimal attack under different variants. They showed circumstances under which selfish miners can hold selfish chain even public chain is longer.

New research areas of selfish mining in machine learning have evolved. Most of these researches have used reinforcement learning methods to improve the optimality of the attack. \cite{RN218, RN221} improved MDP-based solution of \cite{RN216} by applying reinforcement learning algorithms to gain more revenue.\cite{RN223} developed a new deep reinforcement learning framework to analyze the incentives of a rational miner in various conditions and upper bound the security threshold of proof-of-work based blockchain.

\subsection{Learning Automata}
\begin{figure}[!t]
    \centering
    \includegraphics[width=0.5\textwidth]{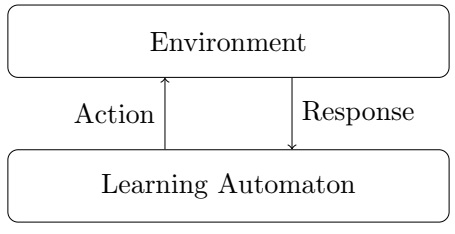}
    \caption{Learning automaton (LA)}
    \label{figure1}
\end{figure}

Learning automata are adaptive decision-making devices that operate in unknown random environments. A Learning automaton has a finite set of actions, and each action has a certain probability (unknown to the automaton) of getting rewarded by its environment. The aim is to learn to choose the optimal action (i.e., the action with the highest probability of being rewarded) through repeated interactions with the system. If the learning algorithm is appropriately chosen, then the iterative process of interacting with the environment can result in selecting the optimal action. The interaction between the learning automaton and the environment is shown in Fig \ref{figure1}.

Learning Automata (LAs) can be classified into two main families, fixed and variable structure learning automata \cite{RN224, RN225, RN227}. Variable action set learning automata used in this paper is a sub-set of variable structure learning automata. Variable action set learning automata can be represented by a sextuple $<\beta,\phi,\alpha,P,G,T>$, where $\beta$ a set of inputs actions, $\phi$ is a set of internal states, $\alpha$ is a set of outputs, P denotes the state probability vector governing the choice of the state at each stage t, G is the output mapping, and T is the learning algorithm. The learning algorithm is a recurrence relation used to modify the state probability vector.

    Such learning automata have a finite set of r actions denoting $<\alpha_1,\alpha_2,...,\alpha_r>$. At each stage t, the action subset $\hat{\alpha}\subseteq{\alpha}$ is available for the learning automata to choose from. Selecting the elements of $\hat{\alpha}$ is made randomly by an external agency. Selecting an action and updating the action probability vector in these automata are as follows.
    Let $S(t)=\sum_{\alpha_i\in\hat{\alpha}(t)}P_i(t)$ presents the sum of probabilities of the available actions in subset $\hat{\alpha}$. Before choosing an action, the available actions probability vector is scaled as Equation \ref{equation1}.
    
\begin{equation}
\label{equation1}
\hat{P}_{i}(t) = \frac{p_i(t)}{S(t)} \quad \forall \alpha_i
\end{equation}

The crucial factor affecting the performance of the variable action set learning automata is the learning algorithm for updating the action probabilities. Let $\alpha_i$ be the action chosen at time t as a sample realization from distribution p(t). Let a and b the reward and penalty parameters, and m denotes the number of available actions. Equations for updating the probability vector are defined by Equation \ref{equation2} for action chosen by learning automata ($i=j$) and Equation \ref{equation3} for other actions($i \neq j$).

\begin{equation}
\label{equation2}
{P}_{j}(n+1) = {P}_{j}(n)+a\beta(1-{P}_{j}(n))-b(1-\beta){P}_{j}(n)
\end{equation}

\begin{equation}
\label{equation3}
{P}_{j}(n+1) = {P}_{j}(n)-a\beta{P}_{j}(n)+b(1-\beta)[\frac{1}{m-1}-{P}_{j}(n)]
\end{equation}

If a=b, the learning algorithm is called the linear reward penalty($L_{RP}$); if $b=\epsilon a$ with $\epsilon < 1$, then the learning algorithm is called linear reward $\epsilon$-penalty ($L_{R_{\epsilon}-P}$), if b=0, the learning algorithm is called the linear reward inaction($L_{RI}$), if a=0, the learning algorithm is called the penalty inaction($L_{PI}$)and finally, if a=b=0, the learning algorithm is called pure chance.
    
Learning automata have found applications in many areas, such as defense mechanisms in network attacks \cite{RN229, RN230}, peer-to-peer networks \cite{RN232}, Internet of Things (IoT) \cite{RN234}, and neural networks \cite{RN236}, to mention a few. In this paper, for the first time, learning automata is used to create a defense mechanism against one of the blockchain attacks.  

\subsection{Properties of an Ideal Defense}
By explaining of problems and weaknesses of existing defenses, as \cite{RN237} suggested, we can enumerate the desirable properties of an ideal defense.
\begin{itemize}
    \item \textbf{Decentralization}: Introducing a trusted server would open a new single point of failure. Moreover, it violates Bitcoin's fundamental philosophy.
    \item \textbf{Incentive Compatibility}: The expected relative revenue of a miner should be proportional to mining power.
    \item \textbf{Backward compatibility}: Non-miners who cannot upgrade their clients can still participate in the network. This is important for hardware products such as Bitcoin ATMs. Specifically, the following rules should not be changed:
    \begin{itemize}
    	\item \textbf{Block validity rules}: A valid block in the current Bitcoin protocol should also be valid within the defense.
    	\item \textbf{Reward distribution policy}: All blocks in the main chain and no other block receive block rewards.
    	\item \textbf{Eventual consensus}: Even when an attack happens, old and new clients should eventually reach a consensus on the main chain.
    \end{itemize}

\end{itemize}

\section{Related work}
In this section, existing defenses are summarized. Summarizing of these defenses is based on the classification of the similarity of methods used in the proposed defenses. The most popular defenses are considered.

Some of the defenses need to make fundamental changes in blockchain structure. Generally, they require significant updates in blockchain nodes, which are incompatible with previous versions. First, Bahack \cite{RN238} proposed a defense with the punishment rule for all miners, including honest miners. In this defense, all miners who fork the blockchain will be punished. The problem with this defense is the punishment of honest miners. Solat et al. \cite{RN239} introduced Zeroblock, where miners are forced to release their blocks within an expected time. If miners withhold their blocks for selfish mining and do not broadcast them within the scheduled time, the peers in the network create their own dummy blocks. These defenses need block validity and reward distribution changes, so network nodes should update their clients to be familiar with the new protocol. \cite{RN241} introduced another defense that will change the structure of transactions. In this defense, the transaction has an extra parameter named Expected Confirmation Height. A comparison of the Expected Confirmation Height and expected value for published block height will use to detect selfish mining attacks. The following paragraph will introduce defenses that operate when a new fork is seen in the network.

Defenses will decrease selfish miners' chances when they create a fork. The most accepted solution against selfish mining is the tie-breaking defense. The tie-breaking defense was proposed by Eyal and Sirer \cite{RN184}. When a miner learns of computing branches of the same length, it should propagate all of them and choose which one to mine on uniformly at random. As they showed in their paper, the minimum mining power needed to start selfish mining will be about 25\%. Heilman \cite{RN242} proposed another backward-compatible defense against selfish mining called Freshness Preferred (FP). In the FP solution, Heilman suggested that each miner use an unforgeable timestamp parameter to penalize miners that withhold blocks by comparing the latest value of the unforgeable timestamp. A trusted party in the network generates an unforgeable timestamp. Heilmen claimed that the lower bound threshold of selfish mining would increase from 25\% to 32\%. The problem with this solution is using a trusted party in the network, which conflicts with the philosophy of Bitcoin's decentralization. The following paragraph will introduce fork-resolving policy-based defenses.

Defenses that will work based on fork-resolving policy. In these defenses, protocols are changed so that when the selfish chain is longer than the public chain, the defense mechanism will work, unlike the tie-breaking defense. The first solution in this category was published by Zhang and Preneel \cite{RN237}, called Publish or Perish. In Publish or Perish, the authors suggested neglecting blocks not published in time and appreciating blocks that incorporate links to competing blocks of their predecessors. Consequently, a block kept secure until a competing block is published contributes to neither or both branches; hence it confers no advantage in winning the block race. The following paragraph will introduce a machine learning-based algorithm to detect selfish mining attacks.

Researches have been done to identify factors that detect selfish mining attacks \cite{RN244, RN246}. These researches use existing data of selfish mining attacks to create training and test data. This research examines factors and tries to find future research areas in this context.

In this paper, the suggested defense algorithm utilizes machine learning to manage parameters and decisions of mining process in a self-adaptive manner. From this perspective, it cannot be compared with all defense mechanisms reported in the literature because there is no machine learning-based defense mechanism in this area. From another perspective, the proposed mechanism suggests a self-adaptive algorithm that is matched with the dynamic and distributed nature of blockchain-based systems. 

\section{Proposed Algorithm: Nik defense}
In this section, the proposed algorithm is presented. At first, the system model in which selfish mining and the proposed defense will work is described. Then, the required definitions for the proposed algorithm are explained. Finally, the proposed algorithm is demonstrated.
\subsection{System Model}
\begin{figure}[!t]
    \centering
    \includegraphics[width=0.5\textwidth]{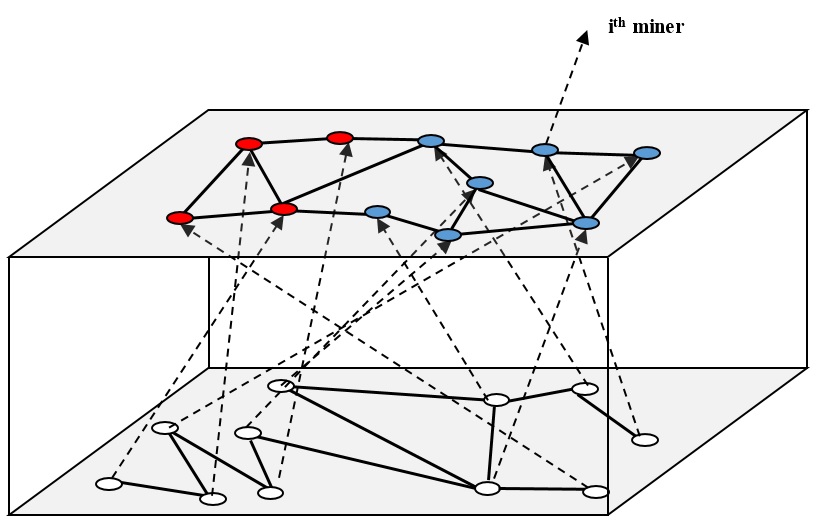}
    \caption{Structure of the miner in the system model}
    \label{figure2}
\end{figure}

In this subsection, a model of the system will be presented. In the proposed model, we consider miner nodes in the blockchain network, so other nodes like super nodes, light nodes, and others will be ignored. Miner nodes in the selfish mining attack form a mining pool to obtain revenue more than their fair share. We will consider two groups of miners to create the most appropriate and potent model. A group of miners follows the selfish mining strategy, which has less than 50\% of total computing power, and a group of miners follows Bitcoin's mining protocol without any deviations. Fig \ref{figure2} shows a snapshot of mining pools in the proposed defense. In this figure, selfish miners which form a mining pool are colored red, and other miners form an honest pool colored blue. The following paragraph will present the distribution of computing power in the network. 

To explain the distribution of computing power in the proposed model, we assume that the selfish mining pool has a $\alpha$ proportion of total computing power and other miners have a $1 - \alpha$ proportion of total computing power. Therefore a newly discovered block with a probability of $\alpha$ belongs to the selfish pool, and a probability of $1 - \alpha$ belongs to other miners. The following paragraph will explain the connectivity of nodes in the proposed model.

We will ignore block propagation delay to clarify network connectivity in our proposed model. This assumption is rational because miners in Bitcoin's network try to send and receive blocks as fast as possible. If they have a delay in the process of sending and receiving, their mining process schedule will disrupt, so they can't find a new block efficiently. Another critical reason to ignore propagation delay is the effort of network researchers and developers to decrease this delay in Bitcoin's network. We see this result in published papers and Bitcoin Improvement Proposals \cite{RN237}. The following paragraph will discuss the arrangement of blocks in the network.

Blocks in every node of Bitcoin form a tree structure. Each block refers to the previous block. For simplicity, in our model, we don't investigate all blocks in every branch of the block's tree. Instead, we can consider only two branches: The main chain or the longest chain, which results from consensus between nodes. Another branch is the private branch which selfish miners have created. None of the honest nodes can't distinguish these two branches. The following paragraphs will discuss the model of miners in the network.

In networks like Bitcoin with a proof-of-work consensus mechanism, we can assume creating a new block as an event in the mining process. On the other hand, creating a new block doesn't relate to time passing, so we can consider the mining process a discrete memoryless model. In this model, every miner, either honest or selfish, decides on the time of finding a new block, and their chosen action continues until the next event finds a new one.

In our proposed model, the selfish mining attacker uses computational power to create a private chain. In arbitrary time t, the selfish mining attacker must choose which block of the main chain to extend for its private chain and which block to release to increase selfish pool revenue.

If the selfish miner realizes that the honest miner finds a new block, it will try to substitute its private block. Parameter $\gamma$ is defined as an advertisement factor. Nodes with $\gamma$ proportion of computing power would accept selfish miner block instead of honest miner block. In terms of block's height, if Bitcoin's network is in $h$ height, the block of the selfish miner in $h$ height with the probability of $\gamma(1-\alpha)$ would accept.Each miner node in the network is equipped with a learning automaton like in Figure \ref{figure2}. Based on i\textsuperscript{th} miner in the network, the required definitions of the proposed algorithm will be explained in the following section.

\subsection{Required Definitions}
To explain the proposed algorithm, the required definitions of the algorithm are defined here. After related definitions, there is an example for explaining definitions in detail.

\textbf{Definition 1.} Competing blocks from i\textsuperscript{th} miner's perspective are defined in this definition. Two blocks compete for being in the main chain if both are valid in obeying protocol rules in creating transactions and blocks and both have the same height. Usually, these blocks form a fork in the block's tree structure.

\textbf{Definition 2.} Weight calculation from i\textsuperscript{th} miner's perspective is defined in this definition. Blocks in competing forks with the same height are compared to calculate the weight of the forks. Between blocks with the same height but in a different fork, the block created recently or has the most timestamp wins the competition, and the related fork's weight will increase one unit. If two blocks have the same timestamp, one of them will be chosen randomly as the competition's winner.

\textbf{Definition 3.} The weight of the fork from i\textsuperscript{th} miner's perspective is defined in this definition. The weight of a fork will be the sum of the assigned weight after calculating the weight of blocks of the same height in different forks. In the proposed method, the fork's weight is denoted by $W$, and the length is denoted by $L$.

\begin{figure}[!t]
    \centering
    \includegraphics[width=0.5\textwidth]{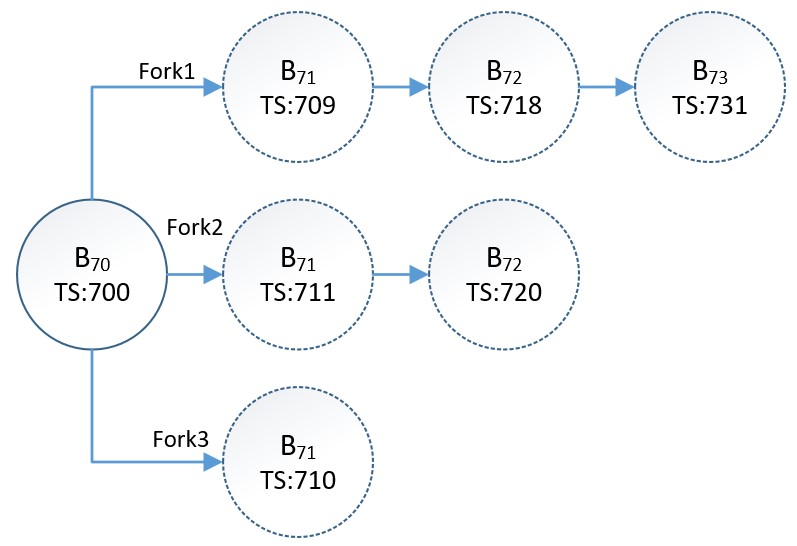}
    \caption{Structure of the block in the system model (TS denotes the timestamp of each blocks)}
    \label{figure3}
\end{figure}

 In Fig \ref{figure3}, an example is designed to illustrate definitions 1, 2, and 3. In the scenario of Figure \ref{figure3}, 70 blocks were mined, and the network nodes had consent about these blocks. At height 71, three forks were created.

    At first, blocks with a height of 71 will be investigated. At this height, fork 2 will win the race, and the weight of fork 2 will increase. Because $B_{71}$ of fork 2 has the most recent timestamp. In height 72, another time, fork 2 will win. At height 73, the only fork with the block is fork 1. So, without any competition, fork 1 will win. By using these definitions, fork 1 has $L=3$, which is the longest chain and fork 2 has $W=2$, which is the heaviest chain.

    There should be a decision between these forks. In the proof-of-work consensus, fork 1 will be chosen because the height of the fork ($L=3$) is longer than the others. But, using the proposed definitions, the weight of forks should be considered in decision making.
    
\textbf{Definition 4.} This definition defines how to choose a fork from i\textsuperscript{th} miner's perspective. Since the miner should choose between different forks that are created by competition of blocks in the same height and maybe one of them is created as a result of the selfish mining attack, need to decide between the length of fork or weight of fork as a base of decision for choosing a fork between different forks. A new parameter is proposed in the proposed method, like [34]. This parameter is called fail-safe and denoted by $K$. When the length of one fork's chain is no longer by $K$ blocks than other forks, the miner should select weight as a base for choosing a chain among forks.

In the Fig \ref{figure3} example, a decision should be made to choose between the weight and height of the chain as a base. If  $K = 1$, the length of fork 1 is longer than the length of fork 2, so the decision is made by Length, and fork 1 will be chosen. On the other hand, if  $K > 1$, the length of fork 1 is no longer than the other by $K$. In this situation, the weight of the chains should be considered, and fork 2 will be chosen.

\textbf{Definition 5.} Fork decision making time from i\textsuperscript{th} miner's perspective is defined in this definition. There should be a time for the miner to check if any forks exist and, if they exist, decide between them. In the proposed method, this time parameter is denoted by $\tau$.

\textbf{Definition 6.} Choosing a fail-safe parameter from i\textsuperscript{th} miner's perspective is defined in this definition. To defend against selfish mining effectively, the miner must change its fail-safe parameter. This change is done using an installed learning automaton on the miner. Learning automaton has three actions (Grow-Shrink-Stop). By selecting one of these actions, the $K$ parameter will change. By evaluating network parameters periodically, one of these actions will be chosen. In the proposed method, this period is called \textit{Time Window}, denoted by $\theta$. $\theta$ is a positive integer coefficient of $\tau$.

\textbf{Definition 7.} Calculating reinforcement signal from i\textsuperscript{th} miner's perspective is defined in this definition. Learning automaton should calculate reinforcement signal, which is denoted by $\beta$. The reinforcement signal is the feedback from the environment, which the learning automaton applied used for updating the learning automaton's probability vector. In the proposed method, this signal is calculated from the analysis of each $\tau$ decision in one $\theta$. Equation \ref{equation4} shows how to calculate $\beta$ after one $\theta$.

\begin{equation}
\label{equation4}
{\beta} = \frac{Number\;of\;Weight\;Decision}{Number\;of\;\{Height\;  +\;Weight \}\;Decision}
\end{equation}

Since the sum of height and weight decisions in a $\theta$ equals the Number of $\tau$, Equation \ref{equation4} can convert to Equation \ref{equation5}.

\begin{equation}
\label{equation5}
{\beta} = \frac{Number\;of\;Weight\;Decision}{Number\;of\;\tau\;in\; \theta}
\end{equation}

\begin{figure}[!t]
    \centering
    \includegraphics[width=0.5\textwidth]{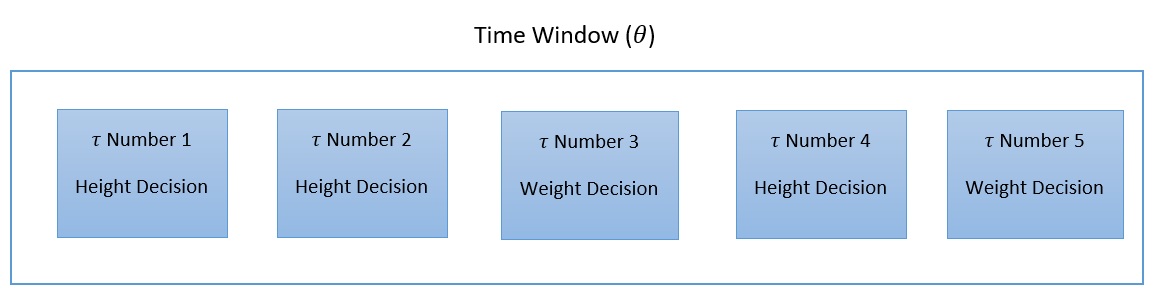}
    \caption{An example of how the miner should calculate reinforcement signal for its learning automaton}
    \label{figure4}
\end{figure}

The following example is designed to illustrate definitions 5, 6, and 7. In this example shown in Figure \ref{figure4}, the defined $\theta$ has $5*\tau$. Based on past decisions, the miner should calculate the reinforcement signal for its learning automaton. Since in $\tau$ number 3, and 5, miners had decisions based on weight. Therefore $\beta$ equals 0.4.

To summarize all notations used in the proposed method, Table \ref{table1} shows the definition for each notation.

\begin{table}[!t]
\renewcommand{\arraystretch}{1.3}
\caption{Table of Notations}
\label{table1}
\centering
\resizebox{\columnwidth}{!}{
\begin{tabular}{cc}
\hline
\bfseries Notation & \bfseries Description\\
\hline\hline
$\alpha$ & The ratio of total computational power that selfish miner has\\
$\gamma$ & The ratio of honest miners that prefer to mine on selfish miners block\\
$W$ & Total weight of chain in the created fork\\
$L$ & Length of chain in the created fork\\
$K$ & Fail-safe parameter for selecting between weight or height as a base for decision \\
$\tau$ & Time parameter for fork checking \\
$\theta$ & The Time Window for changing $\tau$ \\
$\beta$ & Reinforcement signal for miner's learning automata \\
\hline
\end{tabular}}
\end{table}

\subsection{Proposed Algorithm}
After modeling the system and defining the necessary prerequisites for the proposed algorithm, this section will present the proposed algorithm in detail. At first, the proposed algorithm is explained using events that happened while running the defense method; then sub algorithms will be described.
    The proposed algorithm can respond to the occurrence of these events:
    \begin{enumerate}
    \item One
    	Block Receive Event
    	\begin {enumerate}[\IEEEsetlabelwidth{12)}]
    		\item If a fork exists, the miner will check the relation of a new block with existing forks by the previous hash parameter. If the miner needs to create a fork, it will create.
    	\end {enumerate}
    \item 
    	Decision Making Time ($\tau$) Event
    	\begin {enumerate}[\IEEEsetlabelwidth{12)}]
    		\item Existence of a fork will be checked. The miner must select weather by height or weight if a fork exists.
    	\end {enumerate}
    \item 
    Time Window Event
        \begin {enumerate}[\IEEEsetlabelwidth{12)}]
    		\item Existence of a fork will be checked. The miner must select whether by height or weight if a fork exists.
    		\item Reinforcement signal will be created for updating learning automaton.
    		\item Learning automaton will choose the next action. This action uses for updating the $K$ parameter.
    	\end {enumerate}
    \end{enumerate}

In Algorithm \ref{algorithm1}, the proposed algorithm is shown. The proposed algorithm consists of five sub-algorithms. These five sub-algorithms relate to:

\begin{itemize}
\item \textbf{Length Calculation}: Section 4.3.1 explains this algorithm.
\item \textbf{Weight Calculation}: Section 4.3.2 explains this algorithm.
\item \textbf{Chain Selection}: Section 4.3.3 explains this algorithm.
\item \textbf{Action Selection by LA}: Section 4.3.4 explains this algorithm.
\item \textbf{Updating Reinforcement Signal of LA}: Section 4.3.5 explains this algorithm.
\end{itemize}

\begin{algorithm}[H]
\begin{algorithmic}[1]
\renewcommand{\algorithmicrequire}{\textbf{Notation:}}
\renewcommand{\algorithmicensure}{\textbf{Output:}}
\REQUIRE \textit{event} denotes event enum that triggered miner to do action \{TimeWindowEvent, ForkDecisionMakingTimeEvent, BlockReceiveEvent\},\\
{$K_{min}$} denotes the min value for $K$, \\
{$K_{max}$} denotes the max value for $K$, \\
\BEGIN  {1}
\SWITCH {event}
\CASE {TimeWindowEvent}
  \STATE ForkCreationChecking()
  \STATE $\beta$=CalculateUpdateSignal()
  \STATE LA.update($\beta$)
  \STATE UpdateFailSafe($K_{min}$, $K_{max}$)
\ENDCASE
\CASE {ForkDecisionMakingTimeEvent}
  \STATE ForkCreationChecking()
\ENDCASE
\CASE {BlockReceiveEvent}
  \STATE /*Just put the block on the correct fork's chain, and If needed, use ForkSelection algorithm()*/
\ENDCASE
\ENDSWITCH
\END
\end{algorithmic} 
\caption{NikDefense(event)}
\label{algorithm1}
\end{algorithm}

\subsubsection{Length Calculation}
To calculate the length of every chain created by the fork: 1-Getting the height of the last block before the creation of the fork 2-Calculating difference of the last block's height from the height of the last block before the creation of the fork. In Algorithm \ref{algorithm2}, the length calculation algorithm is shown.    

\begin{algorithm}[H]
\begin{algorithmic}[1]
 \renewcommand{\algorithmicrequire}{\textbf{Notation:}}
 \renewcommand{\algorithmicensure}{\textbf{Output:}}
 \REQUIRE {$N$} denotes the number of chains in a fork,\\
{$Chains[i]$ denotes an i\textsuperscript{th} element of chains array in recent $\tau$ time} $i \in [1...N]$,\\
{$ChainsLength[i]$ denotes an i\textsuperscript{th} element of chains array length in the fork} $i \in [1...N]$,\\ 
{$C_H$} denotes the last block's height of the main chain before the creation of the fork,\\
{$LastBlockHeight$} denotes the Height of last block in i
\textsuperscript{th} fork chains \\
\BEGIN {1}
  \FOR {$i\gets1$ to $N$}
  \STATE ChainsLength[i] = Chains[i].LastBlockHeight-$C_H$
  \ENDFOR
 \RETURN ChainsLength
\END
\end{algorithmic}
\caption{ForkChainsLengthCalculation($C_H$)} 
\label{algorithm2}
\end{algorithm}

\subsubsection{Weight Calculation}
To calculate the weight of every chain created by fork: 1-Calculating max length among available chains 2-According to max length, blocks of the different chains but of the same height will be evaluated. Between blocks with the same height, the chain which has the block with the most recent timestamp will win the race 3-Calculation in part 2 will be continued until max length. If one chain is shorter than the others, it will not conclude in comparisons of blocks with higher heights. In Algorithm \ref{algorithm3}, the weight calculation algorithm is shown. 

\begin{algorithm}[H]
\begin{algorithmic}[1]
 \renewcommand{\algorithmicrequire}{\textbf{Notation:}}
 \renewcommand{\algorithmicensure}{\textbf{Output:}}
 \REQUIRE {$N$} denotes the number of chains in a fork,\\
{$Blocks[j][i]$ denotes an i\textsuperscript{th} block of j\textsuperscript{th} chain in recent $\tau$ time} $i \in [1...N]$,\\
{$L_M$} denotes the max length of a fork in a chains array, \\
{$MaxTimestampIndex$} denotes the index of max timestamp in blocks with the same height in different forks \\
\BEGIN {1}
  \FOR {$i\gets1$ to $L_M$}
  \STATE MaxTimestampIndex = 1
  \FOR {$j\gets1$ to $N$}
  \IF {(Blocks[j][i].Timestamp $>$ \\ Blocks[MinTimestampIndex][i].Timestamp)}
  \STATE MaxTimestampIndex = j
  \ENDIF
  \STATE ChainsWeigth[MaxTimestampIndex] += 1
  \ENDFOR
  \ENDFOR
 \RETURN ChainsLength
\END
\end{algorithmic}
\caption{ForksWeightCalculation$(L_M)$} 
\label{algorithm3}
\end{algorithm}

\subsubsection{Chain Selection}
To choose a chain among created chains by fork condition, miner needs to make a decision. So chain selection algorithm of the proposed defense can be described as below: 1-Calculating length of chains 2-Sorting chain based on length in descending order 3-If one chain is longer than the others by $K$, it will select for the next mining event 4-If no chain longer than the others by $K$, the weight of all chains will calculate by the algorithm described in section 4.3.2 5-Sorting chain base on weight in descending order 6-Heaviest chain will be chosen for the next mining event. In Algorithm \ref{algorithm4}, the chain selection algorithm is shown.

\begin{algorithm}[H]
\begin{algorithmic}[1]
 \renewcommand{\algorithmicrequire}{\textbf{Notation:}}
 \renewcommand{\algorithmicensure}{\textbf{Output:}}
 \REQUIRE {$N$} denotes the number of chains in a fork,\\
{$Chains[i]$ denotes an i\textsuperscript{th} element of chains array in recent $\tau$ time} $i \in [1...N]$,\\
{$L_M$} denotes the max length of a fork in a chains array, \\
{$C_H$} denotes the last block's height of the main chain before the creation of the fork,\\
{$ChainsWeight[i]$} denotes an i\textsuperscript{th}chain's weight in a fork, $i \in [1...N]$, \\
{$ChainsLength[i]$ denotes an i\textsuperscript{th} element of chains array length in the fork}, $i \in [1...N]$, \\
\textit{LastBlockHeight} denotes the Height of last block in i\textsuperscript{th} fork chains, \\
\textit{ChosenChain} denotes chosen chain after defense
\BEGIN {1}
  \IF{(N $>$ 1)}
    \STATE ChainsLength = ForkChainsLengthCalculation($C_H$)
    \STATE SortDescendingly(Chains, ChainsLength)
    \IF{(ChainsLength [0] - ChainsLength [1] $>$ K)}
    	\STATE /*Decide based on Chain's Height*/
    	\STATE ChosenChain = Chains[0]
    	\STATE $C_H$ = ChosenChain.LastBlockHeight
    	\RETURN ChosenChain
    \ELSE
    	\STATE /*Decide based on Chain's Weight*/
    	\STATE $L_M$=ChainsLength[0]
    	\STATE ChainsWeight = ChainsWeightCalculation($L_M$)
    	\STATE SortDescendingly(Chains, ChainsWeight)
    	\STATE ChosenChain = Chains[0]
    	\STATE $C_H$ = ChosenChain.LastBlockHeight
    	\RETURN ChosenChain
    \ENDIF 
  \ENDIF
\END
\end{algorithmic}
\caption{ChainSelection()} 
\label{algorithm4}
\end{algorithm}

\subsubsection{Action Selection by LA}
$\theta$ consists of $\tau$ time intervals, and in any of the $\tau$ time intervals particular event can occur; when the $\theta$ has finished, it is needed to make a decision about the occurrence of events in any $\tau$ time intervals of $\theta$. Learning automaton should make this decision as an AI tool. Learning automaton will choose the next $K$ parameter by selecting an action. Chosen action by Learning automaton can be one of these values: 1-\textbf{Grow}: Choosing this action shows that the network was under attack. So Learning automaton increase the $K$ value by one unit to make chain selection harder 2-\textbf{Shrink}: Choosing this action indicates that the network was not under attack or the attack was ineffective. So learning automaton decrease the $K$ value by one unit 3-\textbf{Stop}: Choosing this action by learning automaton shows that the previous chosen action in $\theta$ was correct, and there is no need to change the $K$ value. The reason for selecting variable action set learning automaton is that when the $K$ parameter reaches the max value ($K_{max}$), Grow action should omit from learning automaton's action selection options, and when the $K$ parameter reaches the min value ($K_{min}$), Shrink action should omit from learning automaton's action selection options. In Algorithm \ref{algorithm5}, the action selection algorithm has been shown.

\begin{algorithm}[H]
\begin{algorithmic}[1]
 \renewcommand{\algorithmicrequire}{\textbf{Notation:}}
 \renewcommand{\algorithmicensure}{\textbf{Output:}}
 \REQUIRE {$N$} denotes the number of chains in a fork,\\
{$Chains[i]$ denotes an i\textsuperscript{th} element of chains array in recent $\tau$ time} $i \in [1...N]$,\\
{$ChainsLength[i]$ denotes an i\textsuperscript{th} element of chains array length in the fork} $i \in [1...N]$,\\ 
{$C_H$} denotes the last block's height of the main chain before the creation of the fork,\\
{$LastBlockHeight$} denotes the Height of last block in i
\textsuperscript{th} fork chains \\
\BEGIN {1}
  \IF{($K=K_{max}$)}
    \STATE L = LA.choose \_ action([\text{'}Stop\text{'}, \text{'}Shrink\text{'}])
  \ELSIF{($K=K_{min}$)}
    \STATE L = LA.choose \_ action([\text{'}Grow\text{'}, \text{'}Stop\text{'}])
  \ELSE
    \STATE L = LA.choose \_  action([\text{'}Grow\text{'}, \text{'}Stop\text{'}, \text{'}Shrink\text{'}])
  \ENDIF
\SWITCH {$L$}
\CASE {\text{'}Grow\text{'}}
  \STATE $K = K + 1$
\ENDCASE
\CASE {\text{'}Shrink\text{'}}
  \STATE $K = K - 1$
\ENDCASE
\CASE {\text{'}Stop\text{'}}
  \STATE /*Do nothing about $K$*/
\ENDCASE
\ENDSWITCH
 \RETURN ChainsLength
\END
\end{algorithmic}
\caption{UpdateFailSafe(LA, K, $K_{min}$, $K_{max}$)} 
\label{algorithm5}
\end{algorithm}

\subsubsection{Updating Reinforcement Signal of LA}
By ending $\theta$ and before choosing the next action, learning automaton needs to know how the previous action was by receiving a reinforcement signal. If the previous action was effective, learning automaton would be rewarded, and if the previous action was not effective, learning automaton would be punished. Equation \ref{equation4} shows how the reinforcement signal will help learning automaton to improve itself. In this equation, every $\tau$ interval in $\theta$ is considered. At the end of $\tau$, If the chain is selected by height (it means that one chain is longer than the other by $K$ and does not need to use the weight calculation algorithm), the counter of height selection will increase. Otherwise, the counter of weight calculation will increase.

\section{Proposed Algorithm: SM1 Strategy Modification}
The SM1 strategy is modified in this section to evaluate the proposed defense in the previous section. Since the proposed defense has modified the mining protocol of Bitcoin to create the first AI defense against selfish mining using learning automata, the effort is needed to change the SM1 strategy.
The proposed defense uses $\theta$ and $\tau$ interval parameters, so these parameters should be applied to modify the SM1 strategy. Also, in every $\theta$, the fail-safe parameter will change. To make an effective attack, the attacker should approximate the $K$ parameter using learning automaton. In this paper, approximated $K$ will be shown by \~K. The procedure of approximating \~K  is similar to updating the fail-safe parameter algorithm in Algorithm \ref{algorithm5}. In Algorithm \ref{algorithm6}, the modified version of SM1 is shown.

There are two scenarios in selfish mining attacks that the selfish miner should make a decision about it. Both of these scenarios depend on the length difference between the selfish chain kept by the selfish miner and the honest chain that the honest miner creates. Another factor that can be effective in modified SM1 strategy is approximated \~K parameter.

The first scenario is making a decision on the selfish pool as a decision-maker succeeds in finding a new block. If the private chain and the public chain have one block before finding a new block by the selfish miner, the new block will be added to the selfish chain. In this situation, the selfish chain is ahead, and by publishing its secret blocks, obtaining a reward for two blocks.

The second scenario is making a decision on other miners which considered as honest and succeed in finding a new block. If no private chain is maintained by the selfish node, by receiving an honest block, no action must be taken because the public chain wins the race, and the attack should reset. If the private chain was one block ahead, receiving a new block from an honest node would lose weight calculation due to having a timestamp less than the new received block. So to try its chance, it won't release the private chain. The next block (based on mining by the selfish miner or others) will determine the winner of the game. If the private chain was two blocks ahead, by receiving a new block mined by an honest node, the selfish miner would release its private chain because the new received block wins the first block race in weight calculation by timestamp parameter. In this situation, there is a race between the selfish and honest branches. If the private chain was  \~K+1 was ahead, by receiving a new block mined by an honest node, the selfish miner will release all of the private chain and win \~K+1 blocks revenue. Finally, if a private chain was more than \~K+1 was ahead, by receiving a new block mined by an honest node, the selfish miner will release the first unpublished block.

\section{Evaluation}
This section will describe an evaluation of the Nik defense against selfish mining. To evaluate the proposed defense, we first introduce evaluation metrics. Then, various experiments have been designed to assess the proposed defense and compare it with other defenses. 
\subsection{Metric}
In this subsection, metrics for the evaluation of the proposed defense algorithm are introduced. These metrics will be used in section 6.2 experiments. In the following items, necessary metrics are defined.

\begin{algorithm}[H]
\begin{algorithmic}[1]
\small
 \renewcommand{\algorithmicrequire}{\textbf{Notation:}}
 \renewcommand{\algorithmicensure}{\textbf{Output:}}
 \REQUIRE \textit{$PublicChain$} denotes the public chain which all miners can see, \\
\textit{$PrivateChain$} denotes the chain mined by selfish miners, \\
\textit{$PrivateBranchLength$} denotes the length of the selfish chain, \\
\textit{$\Delta$} denotes the difference of selfish and public chain's length, \\
\textit{\~K} denotes the approximate fail-safe parameter, \\
\textit{$K_{min}$} denotes the min value for $K$, \\
\textit{$K_{max}$} denotes the max value for $K$, \\
\textit{$WeightCalculationCounter$} denotes the number of weight calculations in one ${\theta}$
\BEGIN {1}
  \ON{Init}
  	\STATE PublicChain $\gets$ Publicity Known Blocks
  	\STATE PrivateChain $\gets$ Publicity Known Blocks
  	\STATE PrivateBranchLength = 0
  	\STATE Mine at the head of the private chain
  \ENDON
  \item[]
  \ON{ending ForkDecisionMakingTime}
	\IF{(ReleaseByWeightCalculation() = true)}
		\STATE WeightCalculationCounter = WeightCalculationCounter + 1
	\ENDIF
  \ENDON
  \item[]
  \ON{ending TimeWindow}
  	\STATE $\beta$ = \\CalculateUpdateSignal(WeightCalculationCounter)
  	\STATE LA.update($\beta$)
  	\STATE ApproximateFailSafe(LA,$\widetilde{K}$,$K_{min}$,$K_{max}$)
  	\STATE WeightCalculationCounter = 0
  \ENDON
  \item[]
  \ON{SelfishPool found a block}
    \STATE $\Delta_{prev}$=length(PrivateChain)-length(PublicChain)
  	\STATE Append the new block to PrivateChain
  	\STATE PrivateBranchLength = PrivateBranchLength + 1
	\IF{($\Delta_{prev}=0$ and PrivateBranchLength = 2)}
		\STATE Publish all of the PrivateChain //Try By Luck
		\STATE PrivateBranchLength = 0
	\ENDIF
	\STATE Mine at the head of the PrivateChain
  \ENDON
  \item[]
  \ON{other miners found a block}
    \STATE $\Delta_{prev}$=length(PrivateChain)-length(PublicChain)
  	\STATE Append the new block to PublicChain
	\IF{($\Delta_{prev}$ = 0)}
		\STATE PrivateChain $\gets$ PublicChain //Others win
		\STATE PrivateBranchLength = 0
	\ELSIF {($\Delta_{prev}$ = 1)}
		\STATE /*Do Nothing*/
	\ELSIF {($\Delta_{prev}$ = 2)}
		\STATE Publish all of the PrivateChain //Try By Luck
		\STATE PrivateBranchLength = 0
	\ELSIF {($\Delta_{prev}=\widetilde{K}+1$)}
		\STATE Publish all of the PrivateChain//SelfishPool wins
		\STATE PrivateBranchLength = 0
	\ELSIF {($\Delta_{prev}>\widetilde{K}+1$)}
		\STATE Publish the first unpublished block of the private chain //SelfishPool wins
	\ENDIF
	\STATE Mine at the head of the PrivateChain
  \ENDON
\END
\end{algorithmic}
\caption{Modified SM1()} 
\label{algorithm6}
\end{algorithm}

\begin{itemize}
\item \textbf{Relative Revenue}: This metric is used to measure a miner's revenue based on the revenue of the other miners in the network. Relative revenue is obtained by the number of total accepted mined blocks by one miner divided by the total number of blocks in the main chain. Equation \ref{equation6} is used to show how to calculate the relative revenue of an arbitrary miner.
\small
\begin{equation}
\label{equation6}
\frac{Number\;of\;Mined\;Block\;by\;i\textsuperscript{th}\;Miner}{Total\;Number\;of\;Mined\;Blocks}
\end{equation}
\normalsize

Followed by Equation \ref{equation6}, the relative revenue of the honest miner and the selfish miner can be shown in Equations \ref{equation7} and \ref{equation8}. In these equations, all selfish and honest miners group together and assume that only two groups of miners exist in the network.

\small
\begin{equation}
\scriptsize
\label{equation7}
\frac{Honest\;Miner\;Win\;Block  }{Honest\;Miner\;Win\;Block + Selfish\;Miner\;Win\;Block}
\end{equation}
\normalsize

\small
\begin{equation}
\scriptsize
\label{equation8}
\frac{Selfish\;Miner\;Win\;Block}{Honest\;Miner\;Win\;Block + Selfish\;Miner\;Win\;Block}
\end{equation}
\normalsize

\item \textbf{Lower Bound Threshold}: This metric shows the minimum computing power that selfish miners should provide to start the attack. The lower bound threshold is obtained in simulations by the intersection point of the selfish miner's relative revenue diagram and ideal defense diagram.

\end{itemize}

\subsection{Experiment Results}
Inspiring by previous defenses and their simulators \cite{RN184, RN216, RN237} the proposed algorithm was simulated by converting the mining model into a Monte Carlo simulation process. This conversion makes it possible to distribute newly discovered blocks among selfish and honest miners without solving a cryptographic puzzle. For reproducibility, the developed simulator can be found on Github \footnote{\url{https://github.com/AliNikhalat/SelfishMining}}. Experiments reported in this subsection have been run using Intel Core i7 with a 2.5 GHz frequency clock.
Experiments reported in this section are divided into four parts:
\begin{enumerate}
\item \textbf{Experiment 1}: This experiment aims to compare the proposed defense with the tie-breaking defense as the previous proposed selfish mining defense using a different kind of learning automata. Reward and penalty parameters were changed, and then results were reported. 
\item \textbf{Experiment 2}: This experiment aims to check the effect of changing the K on the proposed defense. By changing the $K$ interval, results of changing this parameter have been reported.
\item \textbf{Experiment 3}: 	This experiment aims to check the effect of changing the $\tau$ time interval on the proposed defense. By changing the $\tau$ time interval, the result of changing this parameter has been reported.
\item \textbf{Experiment 4}: This experiment aims to check the effect of changing the number of $\tau$ in one $\theta$ on the proposed defense. By changing the number of $\tau$, results of changing this parameter have been reported.
\end{enumerate}

For all of the experiments, 10000 blocks were generated by the simulator. The main learning automaton used in these experiments is $L_{R_{\epsilon}-P}$ with a = 0.1, b = 0.01.

\subsubsection{Experiment 1}
\begin{figure}[H]
\centering
\subfloat[Comparison with Tie breaking defense when a=0, b=0]{\includegraphics[width=0.25\textwidth]{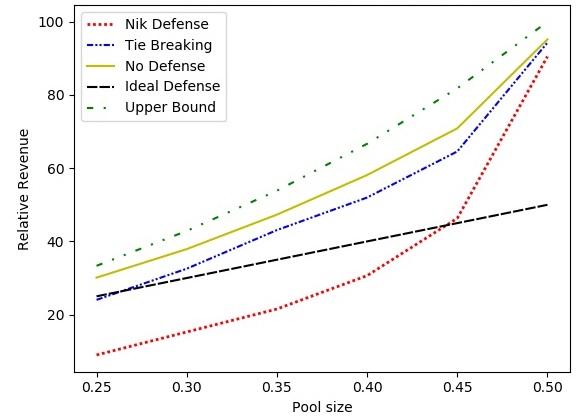}} 
\subfloat[Comparison with Tie breaking defense when a=0.01, b=0]{\includegraphics[width=0.25\textwidth]{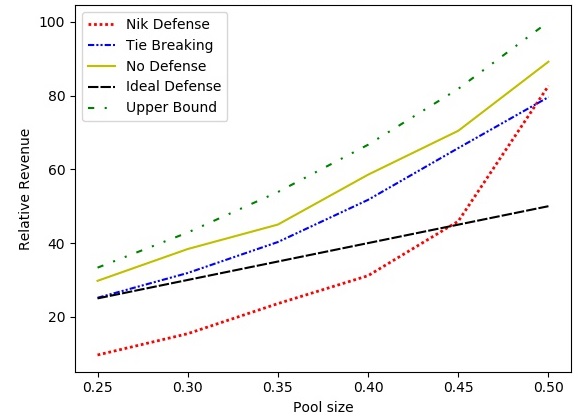}}\\ 
\subfloat[Comparison with Tie breaking defense when a=0, b=0.01]{\includegraphics[width=0.25\textwidth]{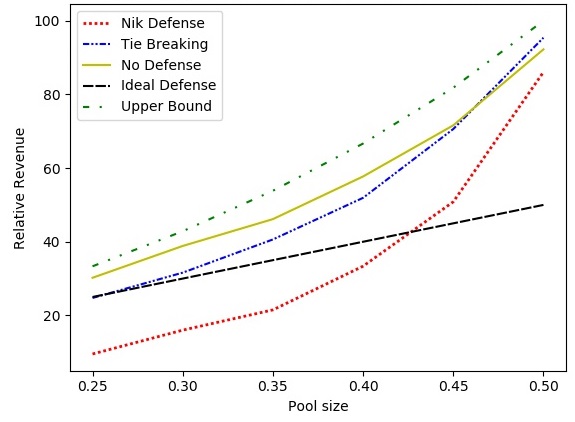}}
\subfloat[Comparison with Tie breaking defense when a=0.01, b=0.01]{\includegraphics[width=0.25\textwidth]{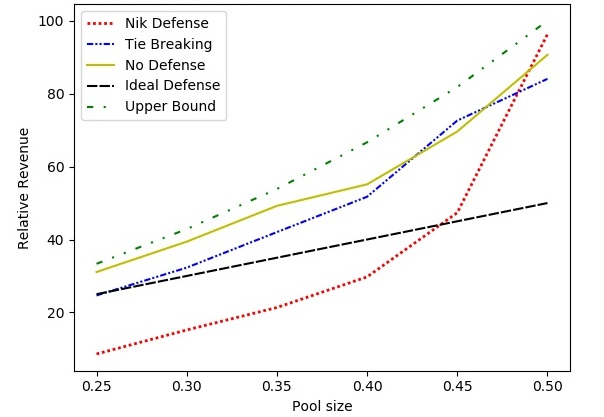}}\\
\subfloat[Comparison with Tie breaking defense when a=0.1, b=0.01]{\includegraphics[width=0.25\textwidth]{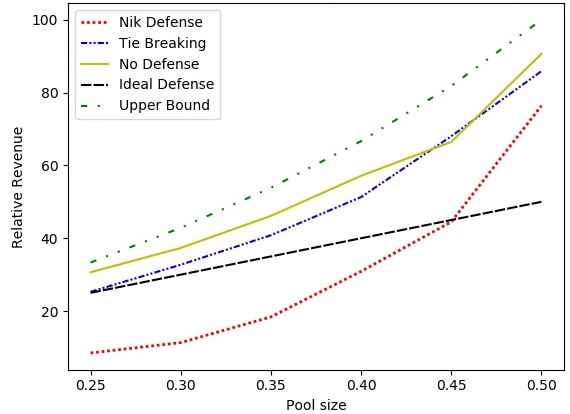}} 
\caption{Comparison with tie-breaking defense using different penalty and reward parameters }
\label{figure5}
\end{figure}

This experiment is conducted to study the impact of changing reward and penalty parameters of learning automaton on the performance of the proposed algorithm with respect to relative revenue. The results of proposed algorithm are compared with the tie-breaking algorithm which is a well-known defense mechanism against selfish mining. For this purpose, parameter $K$ was selected from [1,3] interval. The value of $\tau$ time interval was about the time of mining five blocks, and one $\theta$ has ten $\tau$ time intervals. The results of tie-breaking are compared with those obtained for the proposed defense in five reward and penalty rate values. The results of this experiment are shown in Figure \ref{figure5}.a to Figure \ref{figure5}.e, all of them perform better than the tie-breaking defense. This means that regardless of the type of learning automata, the defense mechanism can prevent the selfish attacker by decreasing the relative revenue. By comparing relative revenue between different learning automata, $L_{R_{\epsilon}-P}$ performs better than the others because the reward rate is more than the penalty rate, and the relative revenue is near the ideal defense. So, choosing the proper reward and penalty rate can affect the quality of defense.

The lower bound threshold value for starting the selfish mining attack is shown in Table \ref{table2}. This parameter can be used to compare starting point of the selfish mining attack's profitability. The lower threshold value is, the less selfish miner attacker is successful. This can be used to show the effectiveness of the defense. According to the results of this experiment, the proposed defense has a higher threshold value than the tie-breaking defense. This threshold increases from 0.25 to 0.4. By increasing the threshold, the selfish miner needs to increase its pool size to start an effective selfish mining attack. 

\begin{table}[!t]
\renewcommand{\arraystretch}{1.6}
\caption{The Threshold for Starting the Selfish Mining Attack}
\label{table2}
\centering
\resizebox{\columnwidth}{!}{
\begin{tabular}{lccccccccc}
\hline
\bfseries & \bfseries  $\boldsymbol P$ & \bfseries $\boldsymbol {L_{R-I}}$ & \bfseries $\boldsymbol {L_{P-I}}$ & \bfseries $\boldsymbol {L_{R-P}}$ & \bfseries $\boldsymbol {L_{R_\varepsilon-p}}$\\
\hline
 \textbf{a} & 0 & 0.01 & 0 & 0.01 & 0.1  \\
 \textbf{b} & 0 & 0 & 0.01 & 0.01 & 0.01  \\
 \hline
 \textbf{Nik} & 0.43 & 0.44 & 0.42 & 0.44 & 0.45  \\
 \textbf{Tie-Breaking} & 0.25 & 0.25 & 0.25 & 0.25 & 0.25  \\
\hline
\end{tabular}}
\end{table}

\subsubsection{Experiment 2}
\begin{figure}[h]
    \centering
    \includegraphics[width=0.5\textwidth]{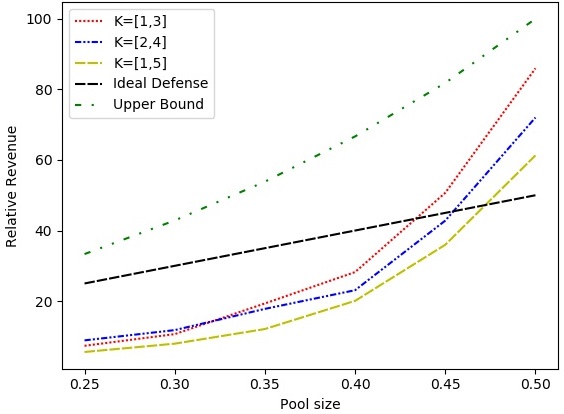}
    \caption{The impact of $K$ intervals on the proposed defense}
    \label{figure6}
\end{figure}

This experiment is conducted to study the impact of the parameter $K$ on the performance of the proposed algorithm with respect to relative revenue. For this purpose, the parameter $K$ was tested for three intervals [1,3], [2,4] and [1,5]. The value of parameter $\tau$ time interval is about the time of mining five blocks, and one $\theta$ has ten $\tau$ time intervals. The results obtained from different values of $K$ are shown in Figure \ref{figure6}. By comparing $K$ intervals, the advantage of choosing bigger values for $K$ is obvious. Figure \ref{figure6} shows [1, 5] interval for $K$ value has the most powerful defense. This can lead to two important factors: 1-learning automaton can choose value 5 for $K$ 2-learning automaton has the freedom to a choose value for $K$ in a bigger interval(choosing four values for $K$). It shows that the interval with bigger values of $K$ and more options for $K$ can reduce the relative revenue of the attacker.

On the other hand, the effect of changing the value of parameter $K$'s on the attack threshold can be seen in Figure \ref{figure6}. [1, 5] has the most threshold for starting an attack. The increasing threshold of starting attack for bigger values of $K$ in the chosen interval can be the result of selecting a bigger $K$ value. Overall, the result shows the bigger values of $K$ leads to the less powerful attack in the network.

\subsubsection{Experiment 3}
\begin{figure}[H]
    \centering
    \includegraphics[width=0.5\textwidth]{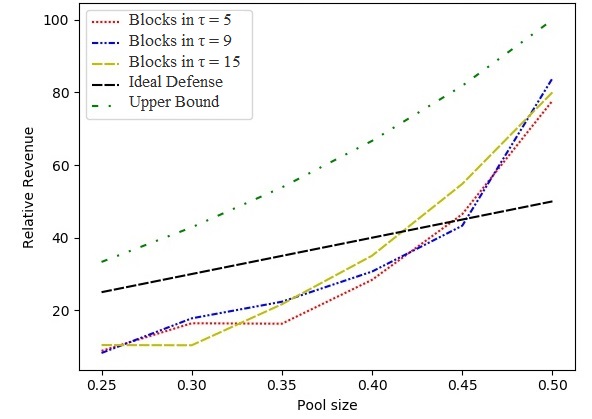}
    \caption{The impact of $\tau$ time interval on the proposed defense}
    \label{figure7}
\end{figure}

This experiment is conducted to study the impact of the $\tau$ time interval parameter on the performance of the proposed algorithm with respect to relative revenue. For this purpose, the parameter $K$ was selected from [1,3] interval. One $\theta$ has ten $\tau$ time intervals. $\tau$ time interval was tested for 5, 9, and 15 mining blocks time. For example, if $\tau=6$, the $\tau$ time interval is about the time of mining six blocks. The results from different values of $\tau$ time intervals are shown in Figure \ref{figure7}. This figure doesn't show  noticeable changes in relative revenue metric after selecting different values for the $\tau$ time interval. This means that small changes in the value of $\tau$ doesn't have a meaningful impact on the performance of the proposed algorithm.

Another important factor shown in Figure \ref{figure7} is the threshold for starting the attack. It shows that a lower value $\tau$ time interval causes an increment in the threshold for starting the attack. As expected, this happened because the lower $\tau$ value in a constant number of $\tau$ (ten time intervals in this experiment) in one  $\theta$ can increase the number of $\theta$ in simulation for the specified number of blocks. Increasing the number of $\theta$ can force the learning automaton to examinate more decisions. By increasing the number of decisions, the learning automaton can tune the $\beta$ parameter effectively, and as a result, the threshold for starting defense will increase. Overall, the result shows a lower value for $\tau$ improves the performance of the proposed defense mechanism.

\subsubsection{Experiment 4}
\begin{figure}[H]
    \centering
    \includegraphics[width=0.5\textwidth]{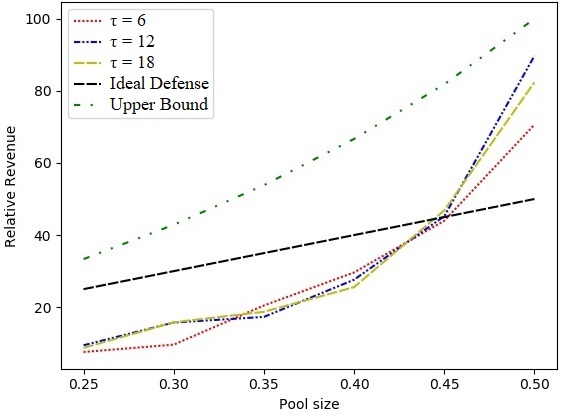}
    \caption{The impact of a number of $\tau$ time intervals in one $\theta$ on the proposed defense}
    \label{figure8}
\end{figure}

This experiment is conducted to study the impact of the number of $\tau$ time intervals parameter in one $\theta$ on the performance of the proposed algorithm with respect to relative revenue. For this purpose, the parameter $K$ was in the [1,3] interval. $\tau$ time interval was about the time of mining five blocks, and one $\theta$ has 6, 12, and 18 $\tau$ time intervals. The results from the different numbers for $\tau$ in one $\theta$ are shown in Figure \ref{figure8}. By comparing the results of this experiment in Figure \ref{figure8}, the lower number of $\tau$ has a better impact on decreasing the relative revenue of selfish miner. As predicted from the behavior of the learning automaton, the less value for $\tau$ in one $\theta$, leads to more numbers of $\theta$ in simulation for the specified number of blocks. Increasing the number of $\theta$ can force the learning automaton to make more decisions. By increasing the number of decisions, the learning automaton can tune the $\beta$ parameter effectively, and as a result, the threshold for starting defense will increase. It means that the lower number of $\tau$ time intervals parameter in one $\theta$ will decrease the effect of the selfish mining attacker in the network.

In addition, Figure \ref{figure8} shows that the number of $\tau$ time intervals creates a little difference in the threshold of starting a selfish mining attack. $\tau=6$ performs better than the others. This happened because of the effective adjusting of the $\beta$ parameter. Eventually, the result shows a lower number of $\tau$ time intervals in one $\theta$ prevents selfish mining efficiently.

\section{Discussion}
Selfish mining attack shows how decentralization of blockchain and Bitcoin as one of the most important implementations of blockchain can be threatened. Considering Bitcoin's popularity, capabilities, and the existence of this threat to Bitcoin, a solution should find to solve this problem. Various solutions have been studied earlier. Existing solutions suffer a lack of self-adaptability. Using learning automata as one of the reinforcement learning tools for designing self-adaptive systems in developing the first AI-based defense, the profitability of selfish mining can decrease.
    We know the problems and limitations of our proposed defense. Our work is designed and implemented in perfect conditions and may not cover all of Bitcoin's protocols, while Bitcoin's network and its protocols are so complicated. In the following, some critical questions have been answered for better discussion.

\begin{itemize}
\item \textbf{Q1: What are the assumptions used in the proposed defense mechanism?}\\
    In the proposed defense mechanism, we assume that every new block has propagated to all nodes before discovering a new block. There are communication delays in sending and receiving messages between nodes. These delays are ignored. This assumption leads to the highest propagation speed to synchronize all nodes as soon as possible.

\item \textbf{Q2: How to set the interval for $\tau$ time interval?}\\
    As examined by experiments performed while writing this paper, time interval has no significant impact on Nik defense before setting $\tau$ time interval. It doesn't mean that time interval is unimportant, but choosing a reasonable interval can help defense mechanism.

\item \textbf{Q3: How to set the number of $\tau$ time intervals in one $\theta$?}\\
    Before setting the number of $\tau$ time intervals, some experiments should be done based on the type of learning automata as we did in this paper. On the other hand, by answering Q1 in this section, communication delays in real situations should be considered in choosing the number of $\tau$ time intervals.   

\item \textbf{Q4: Can different fail-safe parameters collapse the Bitcoin network?}\\
    Choosing a fail-safe parameter can be hard as a synchronization mechanism. Miners may have different values for fail-safe parameters. Our model assumes that all nodes in the network are synced. Different values of the fail-safe parameter in the synced network may lead to a large-scale voting system for choosing the winning branch. At last, miners can choose the winning branch, decreasing the selfish attack's profitability.   
    
\item \textbf{Q5: Why do we use AI for the Nik defense as a self-organized mechanism?}\\
    These days, AI has been increasingly used for detecting and preventing attacks. In blockchain based networks, The events are executed very fast and proposing self-organized and reactive mechanisms that are able to execute appropriate operation after execution of critical events is vital.  Therefore, we have decided to use learning automata based self-adaptive algorithm as an AI tool to prevent selfish mining attacks considering this potential. Blockchain is an unknown space, and since we don't have any information about forks, reinforcement learning of learning automata is a good choice for defense. Hence learning automata were used.  
    
\item \textbf{Q6: What development hurdles can we face while adopting the proposed mechanism?}\\
    The main development hurdle that affect the proposed mechanism is the synchronization algorithms of blocks in the network. The assumption of the upper limit for propagation time of Nik defense can be challenging. It should be noted that this is not limitation of our model and other defense models reported in \cite{RN237, RN238, RN239} face with this problem. 
    
\end{itemize}

\section{Conclusion and Future Research Directions}
The selfish mining attack hurt Bitcoin's incentive compatibility by disrupting the rewarding system's fairness.In addition, the decentralization capability of Bitcoin can be threatened by it. To solve this problem, a new defense based on learning automata as an AI tool was proposed. The proposed method suggested a set of policies to reinforce chain selection algorithm in a self-organized manner. Experiments have shown the superiority of the proposed defense in Bitcoin's network in comparison with existing mechanisms. Compared to existing defenses, the proposed defense can noticeably decrease the profitability of the selfish mining attack. On the other hand, it can increase the threshold for starting the attack's profitability from 25\% to 40\% of the network's computational power. 

Since the proposed study is the first use of an AI-based algorithm for organizing a self-organized defense mechanism, several works in the area of AI to detect and prevent selfish mining attacks can be considered as future directions. In this paper, learning automata theory was used as a self-adaptive decision-maker, but many other models such as Q-learning can be used. This means that other AI tools can modify the proposed algorithm to have a better solution against selfish mining attacks.

In addition, many open problems in the area of proof-of-work consensus need to be studied and explored to develop a more usable and efficient industrial consensus without the selfish mining attack. Alternative consensus algorithms and integration with other systems and architectures are examples of several open issues that can be considered as future directions. This study opens a new horizon for designing defenses mechanism based on AI.

\ifCLASSOPTIONcaptionsoff
  \newpage
\fi



%

%

\begin{IEEEbiography}[{\includegraphics[width=1in,height=1.25in,clip,keepaspectratio]{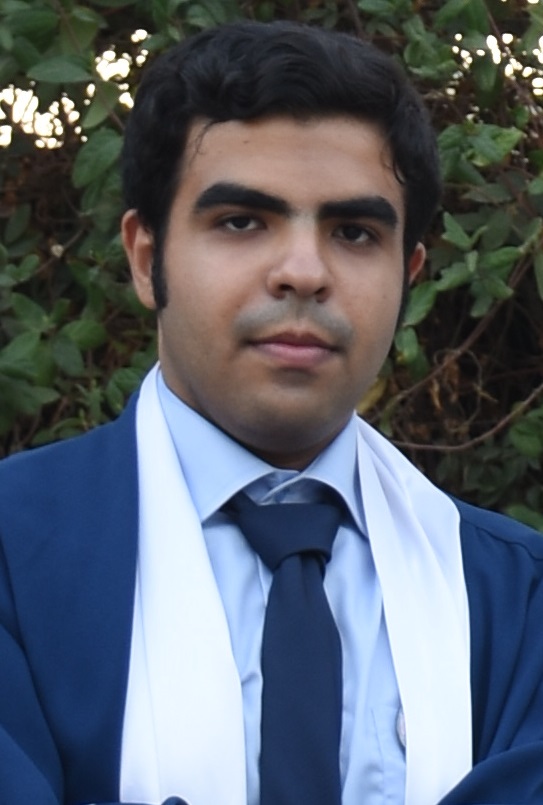}}]{Ali Nikhalat-Jahromi}
He received the B.Sc degree in Electrical Engineering in 2018 and then the M.Sc degree in Computer Engineering in 2021, both from Amirkabir University of Technology, Tehran, Iran. His research interests include Software Systems, Parallel and Distributed Systems, and Programming Languages.
\end{IEEEbiography}

\begin{IEEEbiography}[{\includegraphics[width=1in,height=1.25in,clip,keepaspectratio]{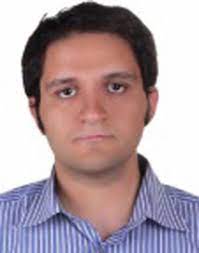}}]{Ali Mohammad Saghiri}
He received the B.Sc. degree from University of Science and Culture, in 2007 and the M.Sc. and Ph.D. degrees from AmirKabir University of Technology, Tehran, Iran, in 2010 and 2017, respectively, all in computer engineering. He published more than 40 scientific papers on international conferences and journals among which Journal of Network and Computer Applications(JNCA), Applied Intelligence, and international journal of communication systems. His research interests include the Internet of Things, Blockchain, and Artificial Intelligence.
\end{IEEEbiography}

\begin{IEEEbiography}[{\includegraphics[width=1in,height=1.25in,clip,keepaspectratio]{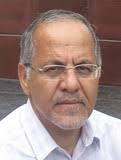}}]{Mohammad Reza Meybodi}
He received the B.Sc. and M.Sc. degrees in economics from Shahid Beheshti University, Tehran, Iran, in 1973 and 1977, respectively, the M.Sc. and Ph.D. degrees in computer science from Oklahoma University, Norman, OK, USA, in 1980 and 1983, respectively.,He was an Assistant Professor with Western Michigan University, Kalamazoo, MI, USA, from 1983 to 1985, and an Associate Professor with Ohio University, Athens, OH, USA, from 1985 to 1991. He is currently a Full Professor with the Computer Engineering Department, Amirkabir University of Technology, Tehran. His research interests include wireless networks, fault tolerant systems, learning systems, parallel algorithms, soft computing, and software development.
\end{IEEEbiography}




\end{document}